\documentclass[aps,prl,reprint,superscriptaddress]{revtex4-1}
\usepackage{blindtext}

\usepackage{xcolor}
\usepackage{hyperref}
\usepackage{epsfig}
\usepackage{subfigure}
\usepackage{graphicx}
\usepackage[toc,page]{appendix}
\usepackage{epstopdf}
\usepackage{amsmath,amssymb,amsfonts}
\usepackage{array}
\usepackage{url}
\usepackage{hyperref}
\usepackage{lineno}
\usepackage{xspace}
\usepackage{listings}
\usepackage{float}
\usepackage{subfigure}

\begin{document}
\title{Beam Energy Dependence of Fifth and Sixth-Order Net-proton Number Fluctuations in Au+Au Collisions at RHIC}
\affiliation{Texas A\&M University, College Station, Texas 77843}
\affiliation{Czech Technical University in Prague, FNSPE, Prague 115 19, Czech Republic}
\affiliation{AGH University of Science and Technology, FPACS, Cracow 30-059, Poland}
\affiliation{Ohio State University, Columbus, Ohio 43210}
\affiliation{Panjab University, Chandigarh 160014, India}
\affiliation{Variable Energy Cyclotron Centre, Kolkata 700064, India}
\affiliation{Brookhaven National Laboratory, Upton, New York 11973}
\affiliation{Abilene Christian University, Abilene, Texas   79699}
\affiliation{Instituto de Alta Investigaci\'on, Universidad de Tarapac\'a, Arica 1000000, Chile}
\affiliation{University of California, Riverside, California 92521}
\affiliation{University of Houston, Houston, Texas 77204}
\affiliation{University of Jammu, Jammu 180001, India}
\affiliation{State University of New York, Stony Brook, New York 11794}
\affiliation{Nuclear Physics Institute of the CAS, Rez 250 68, Czech Republic}
\affiliation{Shanghai Institute of Applied Physics, Chinese Academy of Sciences, Shanghai 201800}
\affiliation{Yale University, New Haven, Connecticut 06520}
\affiliation{University of California, Davis, California 95616}
\affiliation{Lawrence Berkeley National Laboratory, Berkeley, California 94720}
\affiliation{University of California, Los Angeles, California 90095}
\affiliation{Indiana University, Bloomington, Indiana 47408}
\affiliation{Shandong University, Qingdao, Shandong 266237}
\affiliation{Fudan University, Shanghai, 200433 }
\affiliation{Tsinghua University, Beijing 100084}
\affiliation{University of California, Berkeley, California 94720}
\affiliation{ELTE E\"otv\"os Lor\'and University, Budapest, Hungary H-1117}
\affiliation{University of Illinois at Chicago, Chicago, Illinois 60607}
\affiliation{University of Heidelberg, Heidelberg 69120, Germany }
\affiliation{Wayne State University, Detroit, Michigan 48201}
\affiliation{Indian Institute of Science Education and Research (IISER), Berhampur 760010 , India}
\affiliation{Kent State University, Kent, Ohio 44242}
\affiliation{Rice University, Houston, Texas 77251}
\affiliation{University of Tsukuba, Tsukuba, Ibaraki 305-8571, Japan}
\affiliation{Lehigh University, Bethlehem, Pennsylvania 18015}
\affiliation{University of Kentucky, Lexington, Kentucky 40506-0055}
\affiliation{University of Calabria \& INFN-Cosenza, Italy}
\affiliation{National Cheng Kung University, Tainan 70101 }
\affiliation{Purdue University, West Lafayette, Indiana 47907}
\affiliation{Southern Connecticut State University, New Haven, Connecticut 06515}
\affiliation{Central China Normal University, Wuhan, Hubei 430079 }
\affiliation{Technische Universit\"at Darmstadt, Darmstadt 64289, Germany}
\affiliation{Temple University, Philadelphia, Pennsylvania 19122}
\affiliation{Valparaiso University, Valparaiso, Indiana 46383}
\affiliation{Indian Institute of Science Education and Research (IISER) Tirupati, Tirupati 517507, India}
\affiliation{American University of Cairo, New Cairo 11835, New Cairo, Egypt}
\affiliation{Institute of Modern Physics, Chinese Academy of Sciences, Lanzhou, Gansu 730000 }
\affiliation{University of Science and Technology of China, Hefei, Anhui 230026}
\affiliation{Warsaw University of Technology, Warsaw 00-661, Poland}
\affiliation{Frankfurt Institute for Advanced Studies FIAS, Frankfurt 60438, Germany}
\affiliation{National Institute of Science Education and Research, HBNI, Jatni 752050, India}
\affiliation{University of Texas, Austin, Texas 78712}
\affiliation{Rutgers University, Piscataway, New Jersey 08854}
\affiliation{Institute of Nuclear Physics PAN, Cracow 31-342, Poland}
\affiliation{Max-Planck-Institut f\"ur Physik, Munich 80805, Germany}
\affiliation{Creighton University, Omaha, Nebraska 68178}
\affiliation{Indian Institute Technology, Patna, Bihar 801106, India}
\affiliation{Ball State University, Muncie, Indiana, 47306}
\affiliation{Universidade de S\~ao Paulo, S\~ao Paulo, Brazil 05314-970}
\affiliation{Huzhou University, Huzhou, Zhejiang  313000}
\affiliation{Michigan State University, East Lansing, Michigan 48824}
\affiliation{Argonne National Laboratory, Argonne, Illinois 60439}
\affiliation{United States Naval Academy, Annapolis, Maryland 21402}
\affiliation{South China Normal University, Guangzhou, Guangdong 510631}

\author{B.~E.~Aboona}\affiliation{Texas A\&M University, College Station, Texas 77843}
\author{J.~Adam}\affiliation{Czech Technical University in Prague, FNSPE, Prague 115 19, Czech Republic}
\author{L.~Adamczyk}\affiliation{AGH University of Science and Technology, FPACS, Cracow 30-059, Poland}
\author{J.~R.~Adams}\affiliation{Ohio State University, Columbus, Ohio 43210}
\author{I.~Aggarwal}\affiliation{Panjab University, Chandigarh 160014, India}
\author{M.~M.~Aggarwal}\affiliation{Panjab University, Chandigarh 160014, India}
\author{Z.~Ahammed}\affiliation{Variable Energy Cyclotron Centre, Kolkata 700064, India}
\author{D.~M.~Anderson}\affiliation{Texas A\&M University, College Station, Texas 77843}
\author{E.~C.~Aschenauer}\affiliation{Brookhaven National Laboratory, Upton, New York 11973}
\author{J.~Atchison}\affiliation{Abilene Christian University, Abilene, Texas   79699}
% ...

\collaboration{STAR Collaboration}\noaffiliation
\date{February 24, 2023}
%\author{(The STAR Collaboration)}
%\date{July 19, 2022}
%\email{apandav10@gmail.com}
%\author{Myself}
%\author{Someone Else}
%\affiliation{$^1$STAR Collaboration}

\begin{abstract}
We report the beam energy and collision centrality dependence of fifth and sixth order cumulants ($C_{5}$, $C_{6}$) and factorial cumulants ($\kappa_{5}$, $\kappa_{6}$) of net-proton and proton number distributions, from center-of-mass energy ($\sqrt{s_{NN}}$) 3 GeV to 200 GeV Au+Au collisions at RHIC. Cumulant ratios of net-proton (taken as proxy for net-baryon) distributions generally follow the hierarchy expected from QCD thermodynamics, except for the case of collisions at 3 GeV. The measured values of $C_{6}/C_{2}$ for 0-40\% centrality collisions show progressively negative trend with decreasing energy, while it is positive for the lowest energy studied. These observed negative signs are consistent with QCD calculations (for baryon chemical potential, $\mu_{B} \leq$ 110  MeV) which contains the crossover transition range. In addition, for energies above 7.7 GeV, the measured proton $\kappa_{n}$, within uncertainties, does not support the two-component (Poisson$+$Binomial) shape of proton number distributions that would be expected from a first-order phase transition. Taken in combination, the hyper-order proton number fluctuations suggest that the structure of QCD matter at high baryon density, $\mu_{B}\sim 750$ MeV at $\sqrt{s_{NN}}$ = 3 GeV is starkly different from those at vanishing $\mu_{B}\sim 24$ MeV at $\sqrt{s_{NN}}$ = 200 GeV and higher collision energies.
\end{abstract}

\maketitle
An important goal of heavy-ion physics is to study the phase structure of strongly interacting matter. The phase diagram of such strongly-interacting matter, known as the Quantum Chromodynamics (QCD) phase diagram, shows the phase structure as a function of temperature ($T$) and baryon chemical potential ($\mu_{B}$)~\cite{Rajagopal:2000wf,phase_dia_2}.
Lattice QCD (LQCD) calculations have established the  quark-hadron phase transition as a smooth crossover at vanishing $\mu_{B}$~\cite{Aoki:2006we}. At large $\mu_{B}$, QCD-based model calculations indicate that the crossover is replaced by a first-order transition~\cite{Ejiri:2008xt,Bowman:2008kc} which terminates at a critical point. 

Varying the collision energy of heavy nuclei results in a variation in $T$ and $\mu_{B}$ of the strongly-interacting system produced in these collisions, allowing an experimental study of the QCD phase diagram~\cite{Braun-Munzinger:2007edi}. Event-by-event fluctuations or cumulants of net-particle number ($N$) distributions in heavy-ion collisions are sensitive observables for this study~\cite{Stephanov:1999zu,Stephanov:2008qz,Stephanov:2011pb,Asakawa:2009aj}. The cumulants are extensive quantities that can be used to characterize the shape of a distribution. The fifth and sixth-order cumulants, relevant to the current study, are defined as follows: $C_5 =\langle\delta N^5\rangle -10 \langle\delta N^3\rangle \langle\delta N^2\rangle$ and $C_6 = \langle\delta N^6\rangle -15 \langle\delta N^4\rangle \langle\delta N^2\rangle - 10 \langle\delta N^3\rangle^2+30 \langle\delta N^2\rangle^3$, where $\delta N = N - \langle N \rangle$ (For details see Supplemental Material~\cite{supp_BESC56}). For a thermalized system, the ratio of cumulants are directly linked to the susceptibilities ($\chi_{n}$) calculated in a fixed volume, as done in lattice QCD, and in QCD-based and thermal models~\cite{Gavai:2010zn,Gupta:2011wh,Karsch:2010ck,Garg:2013ata}. Experimental measurement of higher order cumulants are also important to understand thermalization in high energy nuclear collisions where the size and duration of the medium is limited~\cite{Gupta:2022phu}. The cumulants, up to the fourth order of various net-particle multiplicity distributions have been analyzed from the first phase of the beam energy scan (BES) program at the Relativistic Heavy-Ion Collider (RHIC) facility~\cite{STAR:2010mib,STAR:2013gus,STAR:2020tga,STAR:2021iop,STAR:2017tfy,STAR:2014egu,STAR:2020ddh,Pandav:2022xxx} and by the HADES experiment at GSI~\cite{HADES:2020wpc}. 
The fourth-to-second order cumulant ratio, $C_{4}/C_{2}$, of net-proton number distributions from the Solenoidal Tracker at RHIC (STAR) experiment shows a non-monotonic collision energy dependence that is qualitatively consistent with expectations from a critical point in the QCD phase diagram~\cite{STAR:2020tga}.

Up to the fourth-order net-proton cumulant ratios, the experimental measurements are positive~\cite{STAR:2020tga} which is reproduced by several model calculations. These include calculations with a crossover quark-hadron transition such as the LQCD~\cite{Bazavov:2020bjn} and the QCD-based functional renormalization group (FRG) model~\cite{Fu:2021oaw}, and those without any phase transition effects like the hadronic transport model UrQMD~\cite{Bleicher:1999xi} and the thermal hadron resonance gas (HRG) model~\cite{Garg:2013ata}. Only after extending the order of fluctuations to five and six (also called hyper-orders) do the theoretical calculations with and without QCD phase transitions show a difference in sign. Negative sign of baryon number susceptibility ratios, $\chi^{B}_{5}/\chi^{B}_{1}$ and $\chi^{B}_{6}/\chi^{B}_{2}$ (also called hyper-skewness and hyper-kurtosis, respectively) is predicted by LQCD~\cite{Borsanyi:2018grb,Bazavov:2020bjn} near the quark-hadron transition temperature for $\mu_{B} \leq$ 110  MeV. The FRG calculations also yield negative $\chi^{B}_{5}/\chi^{B}_{1}$ and $\chi^{B}_{6}/\chi^{B}_{2}$ over a wide $\mu_{B}$ range 24 -- 420 MeV corresponding to central Au+Au collisions at $\sqrt{s_{NN}} = 200 - 7.7$ GeV~\cite{Fu:2021oaw}. Additionally, a particular ordering of susceptibility ratios: $\chi^{B}_{3}/\chi^{B}_{1} > \chi^{B}_{4}/\chi^{B}_{2} > \chi^{B}_{5}/\chi^{B}_{1} >\chi^{B}_{6}/\chi^{B}_{2}$ is predicted by LQCD~\cite{Bazavov:2020bjn}. This is in contrast to the HRG model predictions with an ideal gas equation of state in a grand canonical ensemble framework which remain positive at unity for all ratios~\cite{Borsanyi:2018grb}. 

In search of the first-order phase transition, the factorial cumulants of proton multiplicity distributions have been suggested~\cite{Bzdak:2018axe}. Factorial cumulants, $\kappa_{n}$, up to the sixth order can be defined in terms of cumulants~\cite{Ling:2015yau} as $\kappa_1 = C_1$, $\kappa_2 = -C_1+C_2$, $\kappa_3 = 2C_1-3C_2+C_3$, $\kappa_4 = -6C_1+11C_2-6C_3+C_4$, $\kappa_5 = 24C_1-50C_2+35C_3-10C_4+C_5$ and $\kappa_6 = -120C_1+274C_2-225C_3+85C_4-15C_5+C_6$. The presence of a mixed phase in a first-order phase transition results in a bimodal or two-component structure in the proton multiplicity distribution. Such a bimodal distribution, modeled as Poisson$+$Binomial distributions, yields large factorial cumulants which increase in magnitude and alternate in sign with increasing order~\cite{Bzdak:2018uhv,Bzdak:2018axe}. In probing the two-component nature, the factorial cumulants are less demanding statistically and are more sensitive than regular cumulants~\cite{Bzdak:2018axe}. 

The work reported in this letter is intended to identify the nature of the phase transition over a wide range in $\mu_{B}$ by examining the sign of the hyper-order fluctuations. A recent study of net-proton sixth-order cumulants by STAR hints at a crossover in Au+Au collisions at $\sqrt{s_{NN}} = 200$ GeV ($\mu_{B}  \approx 20$ MeV)~\cite{STAR:2021rls}. In this work, we present new data down to the lowest energy accessible by STAR ($\sqrt{s_{NN}} = 3$ GeV and $\mu_{B} \approx 750$ MeV), along with the measurements of fifth-order net-proton cumulants and fifth- and sixth-order proton factorial cumulants. 

\begin{figure}[!htb]
	\centering 
	\includegraphics[scale=0.355]{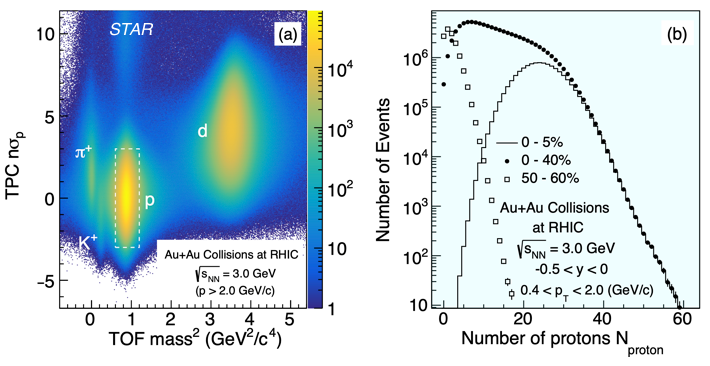}
	\caption{(a) Particle identification using $n\sigma_{p}$ (TPC) versus $m^{2}$ (TOF) for Au+Au minimum bias collisions at 3 GeV (FXT). A momentum criterion $p >$ 2 GeV/$c$ is applied when using $m^{2}$ for proton PID. (b) Proton multiplicity distributions from three collision centralities. These distributions are not corrected for detector efficiency and pile-up effects.}
	\label{fig_3gev_PID_n_DIST} 
\end{figure}

The data from Au+Au collisions having signals in trigger detectors~\cite{Adler:2000bd,Llope:2003ti} above a noise threshold (called minimum bias) at ten collision energies from $\sqrt{s_{NN}} =$ 3 to 200 GeV from the STAR BES-I and fixed-target (FXT) program were analyzed. The number of analyzed events at each energy is summarized in Table~\ref{tab1_stats}.
\begin{table}
	\caption{Total event statistics (in millions) in Au+Au collisions for various collision energies ($\sqrt{s_{NN}}$).}
	\centering   
	\begin{tabular}{|c|c|c|c|c|c|c|c|c|c|c|}
		\hline
		$\sqrt{s_{NN}}$ (GeV) & 3 & 7.7 & 11.5 & 14.5 & 19.6 & 27 & 39 & 54.4 & 62.4 & 200 \\
		\hline
		Events& 140 & 3 & 6.6 & 20 & 15 & 30 & 86 & 550 & 47 & 900 \\
		\hline
	\end{tabular}
	\label{tab1_stats}
\end{table}
The 3 GeV collision data were collected in FXT mode with a constraint on the interaction point (also known as the primary vertex) along the beam axis ($V_z$) of $199.5 < V_{z} < 202$ cm, and the remaining energies were taken in the collider mode of detector operation with $V_z$ within $\pm30$ cm from the center of the STAR detector except for 7.7 GeV data, where $\pm40$ cm was used~\cite{STAR:2021iop,STAR:2017sal}. The tracking and particle identification (PID) are carried out using time projection chamber (TPC) and time of flight (TOF) detectors~\cite{STAR:2002eio}. Protons and antiprotons are required to have rapidity $|y| < 0.5$ at collider energies, and $-0.5 < y < 0$ at 3 GeV due to the asymmetric detector acceptance in the fixed-target mode. The distance of closest approach (DCA) of the (anti-)proton tracks to the primary vertex is required to be less than 1 cm to suppress background~\cite{STAR:2013gus}. The transverse momentum criterion of $0.4< p_{T} <  2.0$ GeV/$c$ is applied at all energies. A variable $n\sigma$~\cite{STAR:2017tfy} that quantifies, in terms of standard deviation, the difference between measured $dE/dx$ from the TPC and its expected value for protons~\cite{Bichsel:2006cs} is utilized for proton identification. We used $|n\sigma| < 2 $. In addition, mass squared ($m^2$) measured using the TOF detector is required to satisfy $0.6 < m^2 < 1.2 $ $\mathrm{GeV}^2/\mathrm{c}^4$ in the $p_{T}$ range $0.8 < p_{T} <  2.0$ GeV/$c$ to achieve high purity for protons~\cite{STAR:2021iop}. For FXT energy at 3 GeV, PID using both TPC and TOF is shown in panel (a) of Fig.~\ref{fig_3gev_PID_n_DIST}. At this energy, if momentum $p \leq 2$ GeV/$c$, only the TPC is used for PID; otherwise, both TPC and TOF are used. The purity of protons in the selected kinematic space is higher than 95\% at all energies~\cite{STAR:2020tga}. Centrality is determined using the charged-particle multiplicity measured by the TPC, excluding protons and anti-protons to avoid self-correlations. Results from 0-40\% and 50-60\% centrality classes are reported. Pile-up events, which happen when separate collisions are reconstructed as a single event, are removed from the analysis by examining the correlation between multiplicities registered in the TPC and TOF~\cite{STAR:2020tga,STAR:2021rls}. Additionally, at higher energies, $\sqrt{s_{NN}} >$ 27 GeV, information from a vertex position detector is used for removing pileup events~\cite{STAR:2021iop}. Because of higher collision rates with the FXT configuration, the pile-up effect becomes large compared to that in collider mode. The correction of cumulants for this effect is then done following the method suggested in Ref.~\cite{Zhang:2021rmu}. 

\begin{figure*}[!htb]
	\centering 
	\includegraphics[scale=0.8]{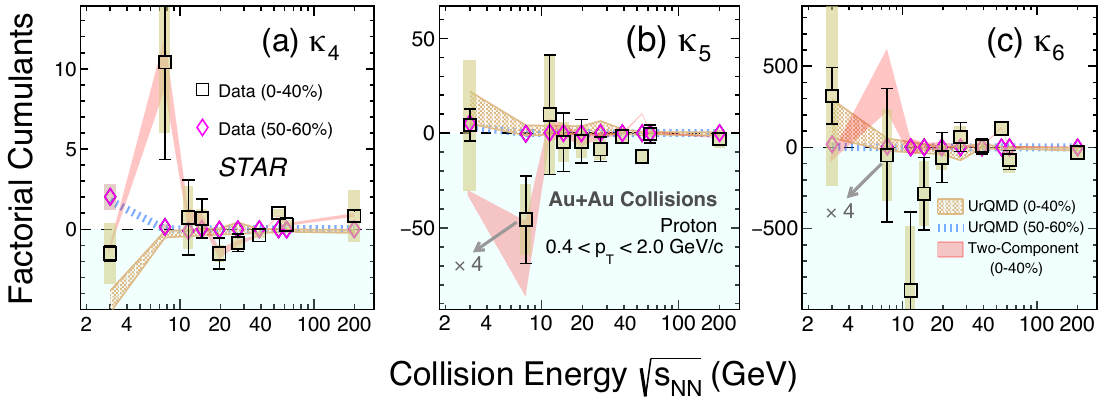}
	\caption{$\kappa_{4}$ (a), $\kappa_{5}$ (b), $\kappa_{6}$ (c) of proton distribution in Au+Au collisions from 3 GeV to 200 GeV. The results are shown for 0-40\% (squares) and 50-60\% (diamonds) centralities. The bars and bands on the data points represent the statistical and systematic uncertainties, respectively. The Two-Component Model (0-40\%) and UrQMD model (0-40\% and 50-60\%) calculations are shown as red, brown bands and blue dashed lines, respectively. The Two-Component Model (with Binomial and Poissonian distributions as constituent components) requires $\kappa_{n}$ up to the fourth order as inputs to predict $\kappa_{5}$ and $\kappa_{6}$. Uncertainties are statistical for the model calculations. The  $\kappa_{5}$ and $\kappa_{6}$ data at 7.7 GeV (0-40\%) are scaled down by a factor of 4 for clarity of presentation.}
	\label{fig_enerdep_K456} 
\end{figure*}

Panel (b) of Fig~\ref{fig_3gev_PID_n_DIST} shows proton multiplicity distributions for 0-5\%, 0-40\% and 50-60\% collision centralities for Au+Au collisions at 3 GeV. Because the number of anti-protons is negligible at this energy (less than the number of protons by 6 orders of magnitude~\cite{kurtosis_3gev}), cumulants of proton distributions are calculated instead of net-proton distributions. Cumulants are then corrected for finite detector efficiency assuming binomial detector response~\cite{Bzdak:2012ab,Kitazawa:2012at,Luo:2014rea,Bzdak:2013pha,Kitazawa:2016awu,Nonaka:2017kko,Luo:2018ofd}. In previous work, relaxing the binomial assumption and implementing an unfolding-based correction for cumulants up to the sixth order for Au+Au collisions at $\sqrt{s_{NN}} = 200$ GeV yielded values consistent with an analytical binomial correction formula within uncertainties~\cite{STAR:2020tga,STAR:2021rls}. To suppress the initial-volume fluctuation effects on cumulants for a given centrality, a centrality bin width correction (CBWC) is performed~\cite{Luo:2013bmi}. While Monte-Carlo studies have shown that at low multiplicities and lower energies residual volume fluctuation effects may remain, the magnitude of the additional correction is highly model dependent~\cite{Sugiura:2019toh,kurtosis_3gev}. Further theoretical understanding of these residual effects are clearly needed before applying to the data and therefore in this analysis only the CBWC is performed. From cumulants, we construct the factorial cumulants and ratios of cumulants which are the observables of this work. The statistical uncertainties on these observables are estimated using the bootstrap method~\cite{Efron:1979bxm,Luo:2014rea,Pandav:2018bdx}. Systematic uncertainties are estimated by varying track selection, particle identification criteria, background estimates (DCA), and track reconstruction efficiency.

Figure~\ref{fig_enerdep_K456} shows collision energy dependence of proton factorial cumulants, $\kappa_{4}$, $\kappa_{5}$ and $\kappa_{6}$  for 0-40\% and 50-60\% centralities. At 7.7 GeV, large positive $\kappa_{4}$ and negative $\kappa_{5}$ are observed for 0-40\% collisions, albeit with large uncertainties. In contrast, at higher energies, the factorial cumulants of all orders show small deviations from zero and from UrQMD expectations. UrQMD calculations reproduce the 3 GeV measurements. The energy dependence trend of the $\kappa_{5}$ and $\kappa_{6}$ measurements is largely reproduced by calculations from a Two-Component Model for proton multiplicity, motivated by the assumption of a first-order phase transition, which inputs in its construction the experimental data of $\kappa_{n}$ up to the fourth order and predicts $\kappa_{5}$ and $\kappa_{6}$~\cite{Bzdak:2018uhv,Bzdak:2018axe} (see Supplemental Material~\cite{supp_BESC56} for details). Vanishing values of factorial cumulants would imply that only the Poissonian part of the Two-Component Model survives.
The small deviation from zero observed for the  proton $\kappa_{n}$ and the absence of a sign change with increasing order for energies above 7.7 GeV within uncertainties does not support the two-component structure for the proton multiplicity distributions at those energies. Note that at 54.4 GeV, a sign change is observed with increasing order for the three factorial cumulants at a level of $2.5-3$ $\sigma_{\rm tot} $ ($\sigma_{\rm tot} $ is the statistical and systematic uncertainties added in quadrature). However the Two-Component Model calculation does not show such a trend. The peripheral 50-60\% measurements are either positive or consistent with zero within uncertainties at all energies.  

\begin{figure*}[!htb]
	\centering 
	\includegraphics[scale=0.8]{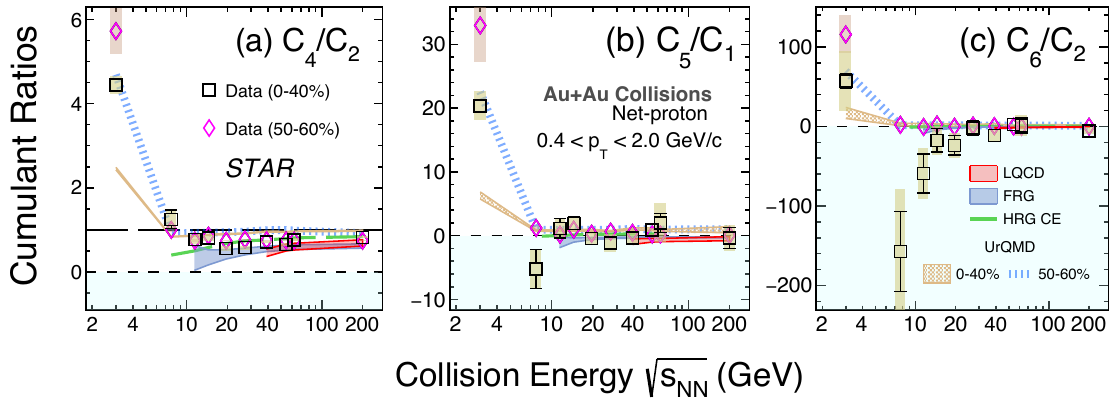}
	\caption{$C_4/C_2$ (a), $C_5/C_1$ (b) and $C_6/C_2$ (c) of the net-proton distribution in Au+Au collisions from 3 GeV to 200 GeV. The results are shown for 0-40\% (squares) and 50-60\% (diamonds) centralities. The bars and bands on the data points represent the statistical and systematic uncertainties, respectively. LQCD ($39 - 200$ GeV)~\cite{Bazavov:2020bjn}, FRG ($11.5 - 200$ GeV)~\cite{Fu:2021oaw}, UrQMD (0-40\%, 50-60\%), and HRG model calculations ($7.7 - 200$ GeV) with canonical ensemble~\cite{Braun-Munzinger:2020jbk} (HRG CE) are shown as red, gray, brown bands, blue and green dashed lines, respectively.}
	\label{fig_enerdep_C456} 
\end{figure*}

\begin{figure*}[!htb]
	\centering 
	\includegraphics[scale=0.7]{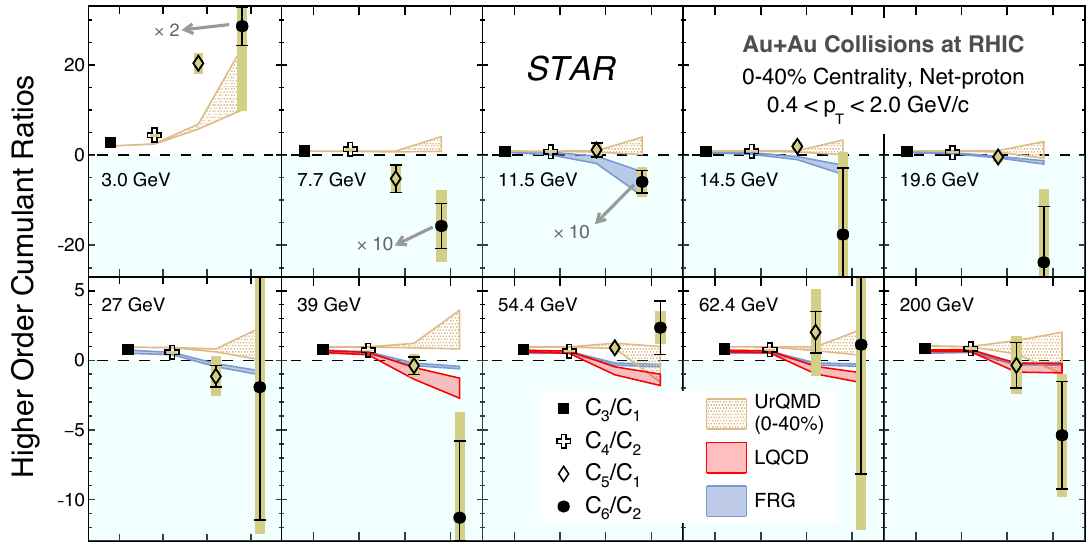}
	\caption{$C_3/C_1$ (filled square), $C_4/C_2$ (open cross), $C_5/C_1$ (open diamond) and $C_6/C_2$ (filled circle) of net-proton distributions in 0-40\% Au+Au collisions from 3 GeV to 200 GeV. The bars and bands on the data points represent the statistical and systematic uncertainties, respectively.  LQCD ($39 - 200$ GeV)~\cite{Bazavov:2020bjn}, FRG ($11.5 - 200$ GeV)~\cite{Fu:2021oaw} and UrQMD calculations (0-40\% centrality) are shown as red, blue and brown bands respectively. The $C_6/C_2$ data at 3 GeV (7.7 and 11.5 GeV) are scaled down by a factor of 2 (10) for clarity of presentation.}
	\label{fig_ordering} 
\end{figure*}

As proxies for net-baryon cumulant ratios~\cite{Kitazawa:2012at}, $C_4/C_2$, $C_5/C_1$ and $C_6/C_2$ of net-proton distributions in Au+Au collisions from 3 GeV to 200 GeV for 0-40\% and 50-60\% centralities are presented in Fig.~\ref{fig_enerdep_C456}. $C_4/C_2$ for 0-40\% centrality is positive at all energies. 
Various model calculations presented for $C_4/C_2$ are also positive. 
$C_5/C_1$ for 0-40\% centrality exhibits weak collision energy dependence and fluctuates about zero with $\lesssim 2.2\sigma_{\rm tot} $ significance except at 3 GeV where it has a large positive value. $C_6/C_2$ for the same centrality is increasingly negative from higher to lower energies down to 7.7 GeV and becomes positive at 3 GeV. The deviations of $C_6/C_2$ from zero at all the energies are within $1.7\sigma_{\rm tot}$. When interpreting the 3 GeV data, one should keep in mind that the initial volume fluctuation effects become significant due to lower charged particle multiplicity.  The increasingly negative sign of $C_6/C_2$ with decreasing energy in the range 7.7 GeV to 200 GeV is qualitatively consistent with LQCD and FRG calculations that include a crossover quark-hadron transition, subject to caveats discussed in Ref.~\cite{STAR:2021rls}. The overall significance of observing negative $C_6/C_2$ in more than half of the collision energies in the range 7.7 GeV to 200 GeV is found to be $1.7\sigma$ (see Supplemental Material~\cite{supp_BESC56}). The UrQMD expectations for these two ratios are either positive or consistent with zero within uncertainties. Expectations from HRG CE are positive for energies greater than 19.6 GeV and become negative only for lower energies (see Supplemental Material~\cite{supp_BESC56} for an enlarged view of model calculations). Recent hydrodynamic calculations also show a similar energy dependence trend as HRG CE~\cite{Vovchenko:2021kxx}. All three ratios are non-negative for peripheral 50-60\% centrality and qualitatively consistent with UrQMD expectations. As the event statistics are lowest at 7.7 GeV (1.2 million events in 0-40\% centrality) among all energies, within the current statistical limitations, the robustness of the negative sign of $C_6/C_2$ at 7.7 GeV (0-40\%) was verified by performing a study on K-statistics~\cite{fisher_kstat} (also known as unbiased estimators of a population's cumulants) and on the sample size dependence of net-proton $C_6/C_2$ which involved creating random samples of varying event statistics from 7.7 GeV data (see Supplemental Material~\cite{supp_BESC56}). Measurements of the three ratios at collider energies using the same rapidity acceptance as for 3 GeV FXT data, i.e., $-0.5< \it{y} <0$, yield similar conclusions regarding the sign as reported here (see Supplemental Material~\cite{supp_BESC56}).

A particular ordering of net-baryon cumulant ratios: $C_3/C_1 > C_4/C_2 > C_5/C_1 > C_6/C_2$, predicted by LQCD was subjected to experimental verification in Fig.~\ref{fig_ordering}. 
Within uncertainties, the measurements for 0-40\% centrality in the energy range 7.7 GeV to 200 GeV are consistent with the ordering expected from LQCD (although at 54.4 and 62.4 GeV, the hierarchy is not as clear as at other energies). While the FRG calculations also follow the predicted hierarchy, the UrQMD calculations within uncertainties do not show any clear ordering and remain non-negative at all energies.
At 3 GeV the cumulant ratios show a reverse ordering: $C_3/C_1 < C_4/C_2 < C_5/C_1 < C_6/C_2$. The probability that the higher energy data would follow a 3 GeV ordering varies between $0.14-10\%$ (see Supplemental Material~\cite{supp_BESC56}). The ordering observed at 3 GeV is reproduced by UrQMD calculations. These observations suggest that the interactions are dominantly hadronic at 3 GeV. Recent results by the STAR experiment on proton $C_4/C_2$ showing suppression at 3 GeV for central 0-5\% Au+Au collisions also supports this inference, indicating that the possible critical point could only exist at collision energies higher than 3 GeV~\cite{kurtosis_3gev}.

In conclusion, measurements of net-proton $C_5/C_1$  and $C_6/C_2$ and proton $\kappa_{5}$ and $\kappa_{6}$ are reported in Au+Au collisions over a broad range of collision energies from 3 GeV to 200 GeV corresponding to a $\mu_{B}$ range of 750 MeV to 24 MeV. The data are presented for 0-40\% and 50-60\% collision centralities. For the first time, we test the ordering of cumulant ratios $C_3/C_1 > C_4/C_2 > C_5/C_1 > C_6/C_2$ expected from QCD thermodynamics. While the overall measured trend for cumulant ratios from 7.7 GeV to 200 GeV seem to follow this hierarchy, a reverse ordering is seen at 3 GeV. $C_6/C_2$ for 0-40\% centrality is increasingly negative with decreasing energy, except at 3 GeV where it is positive. Their deviations from zero at each energy are within $1.7\sigma_{\rm tot}$. The significance of finding negative $C_6/C_2$ (0-40\%) at more than half of the collision energies over the range 7.7 GeV to 200 GeV was found to be $1.7\sigma$. The negative sign of $C_6/C_2$ is consistent with QCD calculations ($\mu_{B} \leq 110$ MeV) that include a crossover quark-hadron transition.
In contrast, the peripheral 50-60\% data, and calculations from the UrQMD model which does not include any QCD transition, are either positive or consistent with zero. 

Proton factorial cumulants $\kappa_{4}$, $\kappa_{5}$, $\kappa_{6}$ (0-40\%) are presented as sensitive observables to probe a possible first-order phase transition~\cite{Bzdak:2018axe}. The measurements indicate the possibility of a sign change at low collision energies, although the uncertainties are large. For energies above 7.7 GeV, the measured proton $\kappa_{n}$ within uncertainties do not support the two-component (Poisson$+$Binomial) shape of proton distributions that is expected from a first-order phase transition.
Peripheral 50-60\% data do not show a sign change with increasing order and are consistent with calculations from the UrQMD model at all energies. The agreement between the presented data and UrQMD at 3 GeV suggests that matter is predominantly hadronic at such low collision energies.
Taken together, the hyper-order proton number fluctuations suggest that the structure of QCD matter at high baryon density, $\mu_{B}\sim 750$ MeV at $\sqrt{s_{NN}} = 3$ GeV is starkly different from those at vanishing $\mu_{B}\sim 24$ MeV at $\sqrt{s_{NN}} = 200$ GeV and higher collision energies. Precision measurements in BES-II with large event statistics will be necessary to confirm these observations. 

We thank the RHIC Operations Group and RCF at BNL, the NERSC Center at LBNL, and the Open Science Grid consortium for providing resources and support.  This work was supported in part by the Office of Nuclear Physics within the U.S. DOE Office of Science, the U.S. National Science Foundation, National Natural Science Foundation of China, Chinese Academy of Science, the Ministry of Science and Technology of China and the Chinese Ministry of Education, the Higher Education Sprout Project by Ministry of Education at NCKU, the National Research Foundation of Korea, Czech Science Foundation and Ministry of Education, Youth and Sports of the Czech Republic, Hungarian National Research, Development and Innovation Office, New National Excellency Programme of the Hungarian Ministry of Human Capacities, Department of Atomic Energy and Department of Science and Technology of the Government of India, the National Science Centre and WUT ID-UB of Poland, the Ministry of Science, Education and Sports of the Republic of Croatia, German Bundesministerium f\"ur Bildung, Wissenschaft, Forschung and Technologie (BMBF), Helmholtz Association, Ministry of Education, Culture, Sports, Science, and Technology (MEXT) and Japan Society for the Promotion of Science (JSPS).

%\blindtext 
%\cite{Pandav:2018bdx}

\bibliographystyle{apsrev4-1} % Tell bibtex which bibliography style to use
\bibliography{references} % Tell bibtex which .bib file to use (this one is some example file in TexLive's file tree)

\section{Supplemental Material}
\subsection{Two-Component Model calculations}
In a two-component, or bimodal, distribution, the total probability distribution $P(N)$ is a combination of two separate constituent distributions, $P_A(N)$ and $P_B(N)$, so that
\begin{eqnarray}
P(N)= (1-\alpha)P_A(N)+\alpha P_B(N),
\label{eqn_bimodal}
\end{eqnarray}
where the parameter $\alpha$ ( $\alpha \leq 1$) specifies the relative contribution of the two.
The factorial cumulants ($\kappa_n$) of such a distributions up to the sixth order can be expressed in terms of factorial cumulants of the two constituent distributions ($\kappa_{nA}$ and $\kappa_{nB}$) as follows~\cite{Bzdak:2018uhv}.
\begin{widetext}
\begin{eqnarray}
%\kappa_1 &=& (1-\alpha)\kappa_{1A}+\alpha \kappa_{1B}\\
\kappa_1 &=& \kappa_{1A}-\alpha \Delta \kappa_1\\
\kappa_2 &=& \kappa_{2A}-\alpha [\Delta \kappa_2-(1-\alpha)\Delta \kappa_1^2]\\
\kappa_3 &=& \kappa_{3A}-\alpha [\Delta \kappa_3(1-\alpha)((1-2\alpha) \Delta \kappa_1^3-3 \Delta \kappa_1 \Delta \kappa_2)]\\
\kappa_4 &=& \kappa_{4A}-\alpha[ \Delta \kappa_4-(1-\alpha)((1-6\alpha+6\alpha^2)\Delta \kappa_1^4-6(1-2\alpha)\Delta \kappa_1^2\Delta \kappa_2\nonumber\\
&+&4\Delta \kappa_1\Delta \kappa_3+3\Delta \kappa_2^2)]\\
\kappa_5 &=& \kappa_{5A}-\alpha[ \Delta \kappa_5+(1-\alpha)((1-2\alpha)(1-12\alpha+12\alpha^2)\Delta \kappa_1^5\nonumber\\
&-&10(1-6\alpha+6\alpha^2)\Delta \kappa_1^3 \Delta \kappa_2+10(1-2\alpha)\Delta \kappa_1^2 \Delta \kappa_3\nonumber\\
&+&15(1-2\alpha) \Delta \kappa_1\Delta \kappa_2^2-5\Delta \kappa_1 \Delta \kappa_4-10\Delta \kappa_2 \Delta \kappa_3)]\\
\kappa_6 &=&  \kappa_{6A}-\alpha[\Delta \kappa_6-(1-\alpha)((1-30\alpha(1-\alpha)(1-2\alpha)^2)\Delta \kappa_1^6\nonumber\\
&-&15(1-2\alpha)(1-12\alpha+12\alpha^2)\Delta \kappa_1^4\Delta \kappa_2+20(1-6\alpha+6\alpha^2)\Delta \kappa_1^3\Delta \kappa_3\nonumber\\
&-&15\Delta \kappa_1^2(\Delta \kappa_4(1-2\alpha)-3\Delta \kappa_2^2(1-6\alpha+6\alpha^2))\nonumber\\
&+&6\Delta \kappa_1(\Delta \kappa_5-10\Delta \kappa_2\Delta \kappa_3(1-2\alpha))\nonumber\\
&-&15(1-2\alpha)\Delta \kappa_2^3+10\Delta \kappa_3^2+15\Delta \kappa_2\Delta \kappa_4)]
\label{eqn_bimodal_facto}
\end{eqnarray}
\end{widetext}
where $\Delta \kappa_n$ = ($\kappa_{nA} - \kappa_{nB}$) for $n$ = 1, 2, 3, 4, 5, 6.

%To obtain the Two-Component Model expectations for fifth and sixth order proton factorial cumulants, we follow the procedure as suggested in
For the Two-Component Model used in this letter to calculate expectations for the fifth- and sixth-order proton factorial cumulants, we follow the procedure suggested in Refs.~\cite{Bzdak:2018uhv,Bzdak:2018axe}. The two constituent distributions of this Two-Component Model are the binomial ($P_A(N)$) and Poissonian ($P_B(N)$) distributions; this choice is made keeping in mind the baryon number conservation~\cite{Bzdak:2018uhv}. The binomial distribution has two parameters: number of trials ($B$) and probability of success ($p$), while the Poissonian has only one parameter, the mean ($\lambda$). Thus, a Two-Component Distribution with binomial and Poisson distributions as constituents has four parameters in total: $\alpha$, $B$, $p$ and $\lambda$. Factorial cumulants of the binomial and Poisonian distributions can be deduced from their parameters. Following the recommendation in Ref.~\cite{Bzdak:2018uhv}, we fix the value $B=350$ and then using data and the equations for the first-, third-, and fourth-order factorial cumulants, we extract the remaining parameters. Note that at $\sqrt{s_{NN}}$ = 7.7 GeV, the equations for first-, second-, and fourth-order factorial cumulants were employed for extracting the parameters, as the former choice of equation resulted in unphysical values of the $\alpha$ parameter ($\alpha>$1). Nonetheless, both sets of parameters predict the sign change and comparable values of fifth- and sixth-order factorial cumulants. The extracted parameters at all energies are summarized in Table~\ref{tab1_param}. With all four parameters of the Two-Component Distributions known, predictions are made for the fifth- and sixth-order factorial cumulants. To evaluate the statistical uncertainties on the predictions, the resampling method suggested in Ref.~\cite{Bzdak:2018axe} is performed.

\begin{table*}
	\caption{The parameters of the Two-Component Model calculations at all collision energies.}
	\centering   
	\begin{tabular}{|c|c|c|c|c|c|c|c|c|c|c|}
		\hline
		$\sqrt{s_{NN}}$ (GeV) & 3 & 7.7 & 11.5 & 14.5 & 19.6 & 27 & 39 & 54.4 & 62.4 & 200 \\
		\hline
		$p$& 0.04035 & 0.0601 & 0.0483 & 0.0411 & 0.0429 & 0.03088 & 0.0284 & 0.0647 & 0.0285 & 0.0313 \\
		\hline
		$\alpha$& 0.306096 & 0.00745336 & 0.0642206 & 0.0221282 & 0.793785 & 0.562374 & 0.587115 & 0.999963 & 0.0226576 & 0.0135979 \\
		\hline
		$\lambda$& 16.98 & 15.16 & 18.32 & 16.46 & 12.78 & 12.39 & 11.22 & 10.17 & 11.73 & 8.3 \\
		\hline
	\end{tabular}
	\label{tab1_param}
\end{table*}

\subsection{Cumulants and $K$-statistics}
Cumulants quantify the characteristics of a distribution. The cumulants of a distribution up to the sixth order are defined as
\begin{eqnarray}
%\begin{allign*}
C_1 &=& m_1\\
C_2 &=& \mu_{2}\\
C_3 &=& \mu_{3}\\
C_4 &=&\mu_{4}-3\mu_{2}^2\\
C_5 &=&\mu_{5}-10\mu_{3}\mu_{2}\\
C_6 &=&\mu_{6}-15\mu_{4}\mu_{2}-10\mu_{3}^2+30\mu_{2}^3
\label{eqn_cumu}
\end{eqnarray}
where $m_1$ is the first raw moment or the mean and $\mu_n$ ($n \geq 2$) are the central moments defined as $\mu_n$ =  $\langle(\delta N^n)\rangle$ with $N$ being the variate whose distribution is considered and $\delta N$ = $N$ - $m_1$. The cumulants are well known statistical quantities. For example, the first- and second-order cumulants are the mean and variance. The third- and fourth-order cumulants reflect the skewness and kurtosis of a distribution, respectively.
\\
\\
In most statistical analyses, the information about the population is not known a priori but rather inferred using the sample. One only has a given sample to work with, which forms a subset of the population. Measurements are performed on the available sample to infer the traits of the population. If the sample size is sufficiently large, cumulant measurements of the sample itself can serve as reasonable estimates of a population's cumulants. $K$-statistics are known to be unbiased estimators of a population's cumulants~\cite{fisher_kstat}. If a sample size is sufficiently large, the $K$-statistics and cumulants of the sample should be consistent with one another, so measuring both and comparing them is one method to assess the adequacy of a sample size. $K$-statistics ($KS_n$) up to sixth-order can be expressed in terms of central moments ($\mu_n$) of the sample as follows~\cite{fisher_kstat}. 

\begin{figure}[!htb]
	\centering 
\includegraphics[scale=0.8]{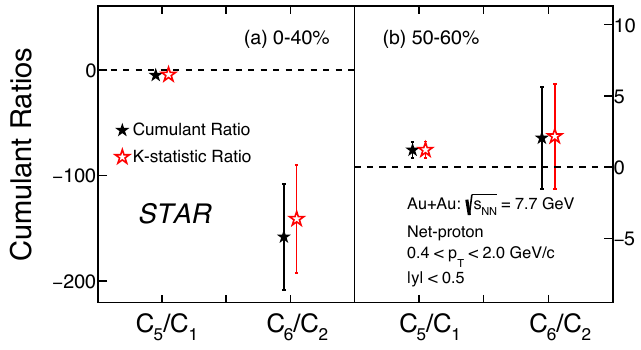}
\caption{Net-proton cumulant ratio $C_5/C_1$ and $C_6/C_2$ and their corresponding $K$-statistics ratios in Au+Au collisions at  $\sqrt{s_{NN}}$= 7.7 GeV in 0-40\% (a) and 50-60\% (b) collision centralities. Only statistical uncertainties are shown.}
\label{kstat_cen5} 
\end{figure}

\begin{widetext}
\begin{eqnarray}
%\begin{allign*}
KS_1 &=& C_1\\
KS_2 &=& \frac{n}{n-1}\mu_{2}\\
KS_3 &=& \frac{n^2}{(n-1)(n-2)}\mu_{3}\\
KS_4 &=& \frac{n^2}{(n-1)(n-2)(n-3)}[(n+1)\mu_{4}-3(n-1)\mu_{2}^2]\\
KS_5 &=& \frac{n^3}{(n-1)(n-2)(n-3)(n-4)}[(n+5)\mu_{5}-10(n-1)\mu_{2}\mu_{3}]\\
KS_6 &=& \frac{n^2}{(n-1)(n-2)(n-3)(n-4)(n-5)}[(n+1)(n^2+15n-4)\mu_{6}\nonumber\\&-&15(n-1)^2(n+4)\mu_{2}\mu_{4}-10(n-1)(n^2-n+4)\mu_{3}^2+30n(n-1)(n-2)\mu_{2}^3]
\label{eqn_kstat}
\end{eqnarray}
\end{widetext}
When the sample size ($n$) is large such that $n\sim(n-1)$, the $K$-statistics and cumulants converge to the same value as can be clearly seen in the example of $KS_2$. Among all the STAR Au+Au data samples, $\sqrt{s_{NN}}$ = 7.7 GeV has the fewest number of recorded events. We calculated the ratio of fifth-to-first and sixth-to-second order $K$-statistics of net-proton distributions in Au+Au collisions at $\sqrt{s_{NN}}$ = 7.7 GeV for 0-40\% and 50-60\% collision centralities, and their comparison with cumulant ratios $C_5/C_1$ and $C_6/C_2$ is shown in Fig.~\ref{kstat_cen5}. These $K$-statistics ratios of fifth and sixth orders are consistent with the corresponding cumulant ratios, which demonstrates that the statistics in $\sqrt{s_{NN}}$ = 7.7 GeV are sufficient for the fifth- and sixth-order K-statistics and cumulants to agree.

\subsection{Statistics dependence of $C_6$/$C_2$}
\begin{figure*}[!htb]
	\centering 
\includegraphics[scale=0.7]{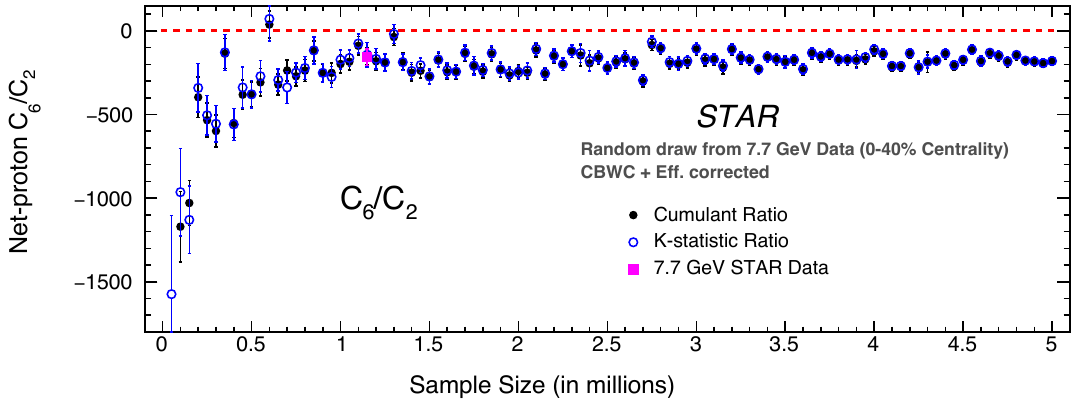}
\caption{Net-proton cumulant ratio $C_6/C_2$ (filled black circles) and the corresponding $K$-statistics ratio (open blue circles) as a function of subsample size. Samples of different sizes are created by random draws from STAR Au+Au collision data at $\sqrt{s_{NN}}$ = 7.7 GeV, 0-40\% centrality. Measurements are centrality bin width corrected (CBWC) and also corrected for efficiency. The STAR data for 0-40\% centrality (filled magenta square) is also shown. Only statistical uncertainties are shown. }
\label{kstat_depen_c62} 
\end{figure*}

We performed random sampling from the STAR data to create sub-samples of various sizes in order to study sample size dependence of sixth-order cumulants and $K$-statistics. Events were randomly drawn with replacement to create these sub-samples, with each event having the same probability of being chosen. As higher order cumulants are known to be statistics-hungry, we conducted this study for the sixth-order cumulant using STAR data at $\sqrt{s_{NN}}$ =7.7 GeV (because it is the smallest sample), and the net-proton $C_6/C_2$ was observed to have a large negative value for 0-40\% centrality. Figure ~\ref{kstat_depen_c62} shows net-proton $C_6/C_2$ as a function of sample size where each of the simulated samples (referred to below as "subsamples") are independently drawn from observed events at $\sqrt{s_{NN}}$ =  7.7 GeV (0-40\% centrality). Also shown are the $K$-statistics ratio measurements of the same order. All the necessary corrections done with STAR data are also carried out for each subsample. The 7.7 GeV data set from STAR has $\sim$1.2 million events in 0-40\% centrality. In this study, subsample size is varied in the range of 0.05 - 5 million events, in steps of 0.05 million. In the entire range of subsample size studied, the net-proton $C_6/C_2$ is negative (with the exception of a very few cases where is it consistent with zero within uncertainty). The value is more negative in the smaller-statistics subsamples and as the subsample size increases, the value saturates near the observed value in the true data sample. No cases with positive $C_6/C_2$ were found. The $K$-statistics ratio of the same order also shows a similar trend and remains consistent with the cumulant ratio $C_6/C_2$ within uncertainties.
One caveat in this study to be kept in mind is that while performing random draws from the real data, the subsample events will be constrained by the total available events in STAR data. 

\subsection{Model calculations for Net-proton $C_4/C_2$, $C_5/C_1$ and $C_6/C_2$  }
\begin{figure*}[!htb]
	\centering 
\includegraphics[scale=0.7]{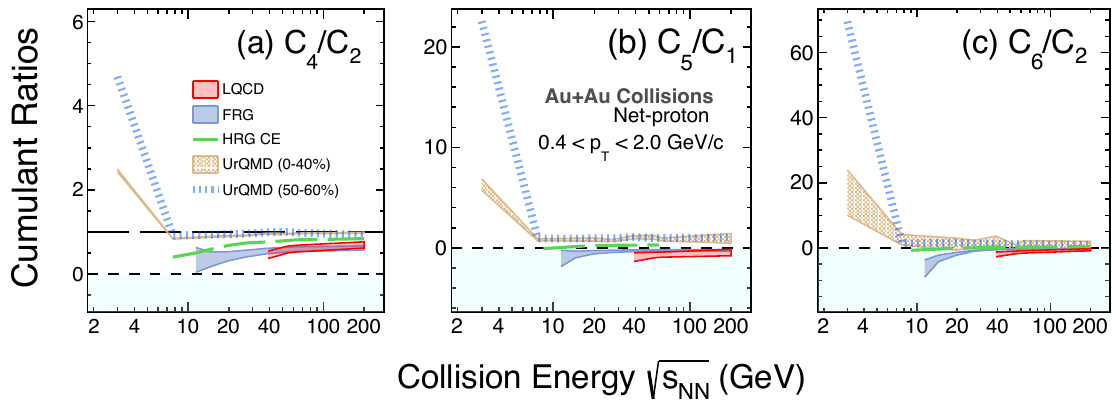}
\caption{Same as Fig. 3 with only model results. Lattice QCD (39 -- 200 GeV)~\cite{Bazavov:2020bjn}, FRG (11.5 -- 200 GeV)~\cite{Fu:2021oaw}, UrQMD (0-40\%, 50-60\%)~\cite{Bleicher:1999xi}, and HRG model calculations (7.7 -- 200 GeV) with canonical ensemble~\cite{Braun-Munzinger:2020jbk} are shown as red, grey, brown bands, blue and green dashed lines, respectively.}
\label{ratios_model} 
\end{figure*}
Figure~\ref{ratios_model} reproduces an enlarged version of the model calculations already presented in Fig. 3. The lattice QCD (LQCD) and functional renormalization group (FRG) model predictions are negative for $C_5/C_1$ and $C_6/C_2$ while they are positive for $C_4/C_2$. The UrQMD model calculations are either positive or consistent with zero within uncertainties for the two centralities presented. The HRG model calculations with canonical ensemble (HRG CE) yield positive values of the three net-proton cumulant ratios except at low collision energies, where $C_5/C_1 <$  0  ($\sqrt{s_{NN}} <$   11.5 GeV) and $C_6/C_2 <$  0  ($\sqrt{s_{NN}} <$  19.6 GeV).

\subsection{$C_5$ and $C_6$ measurements in rapidity range -0.5$<y<$0}
\begin{figure*}[!htb]
	\centering 
\includegraphics[scale=0.7]{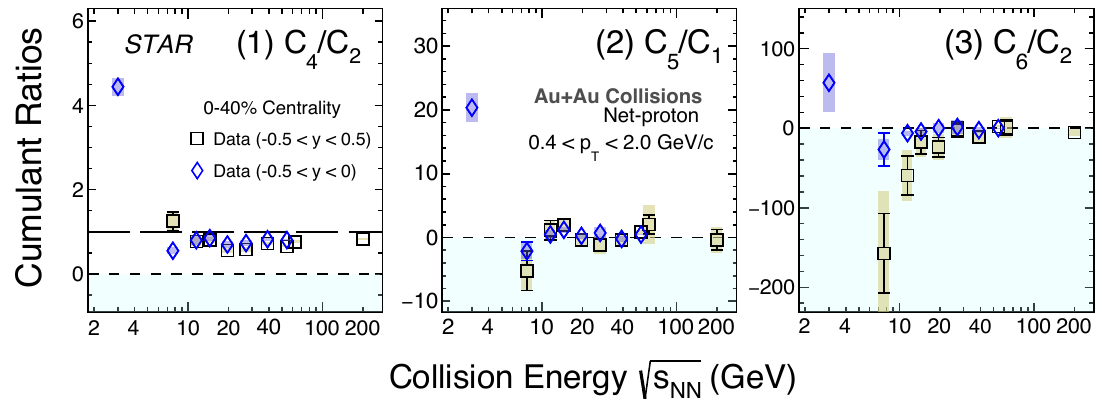}
\caption{Same as Fig. 3 except that the net-proton $C_4/C_2$ (a), $C_5/C_1$ (b) and $C_6/C_2$ (c) measurements with rapidity window $-0.5 < y < 0$ for (anti-)proton selection for 0-40\% centrality are also shown for Au+Au collisions from $\sqrt{s_{NN}}$ = 7.7 -- 54.4 GeV (blue diamond markers). The bars and bands on the datapoints represent statistical and systematic uncertainties, respectively.}
\label{half_eta_cover} 
\end{figure*}

The fifth- and sixth-order fluctuations reported in this work spans a broad range of energies for Au+Au collisions, from $\sqrt{s_{NN}}$ =  3 -- 200 GeV. The key aspect of the measurement is to look for a sign change in fifth- and sixth-order cumulants. While the Au+Au collision dataset collected in the collider mode of detector operation allows for symmetric rapidity acceptance $-0.5 < y < 0.5$ for (anti-)proton selection, the detector acceptance in the FXT mode forbids such a choice of rapidity window and thus the measurements instead were carried out with (anti-)proton rapidity range of $-0.5 < y < 0$. As a check for any systematic effect of this difference in rapidity acceptance, the net-proton $C_4/C_2$, $C_5/C_1$ and $C_6/C_2$  with (anti-)proton selected within -0.5$<y<$0 were also measured in the collider energies from $\sqrt{s_{NN}}$ =  7.7 -- 54.4 GeV for the 0-40\% centrality. As shown in Fig.~\ref{half_eta_cover}, the sign and the energy dependence of the net-proton $C_5/C_1$ and $C_6/C_2$ measured with $-0.5 < y < 0$, are largely consistent with those obtained with the rapidity range $-0.5 < y < 0.5$ (default case). The sign of $C_4/C_2$ for the two rapidity windows are also consistent, i.e. positive at all collision energies presented. The three cumulant ratios at collider energies are seen to be closer to the Poisson limit at unity when the rapidity coverage of measurements is decreased from $-0.5 < y < 0.5$ to $-0.5 < y < 0$.

\subsection{Significance calculation}
\paragraph*{{\bf{Negative sign of net-proton $C_6/C_2$ from $\sqrt{s_{NN}}$ =  7.7 -- 200 GeV:}}}
The net-proton $C_6/C_2$ for 0-40\% centrality in Au+Au collisions from nine collision energies in the range $\sqrt{s_{NN}}$ =  7.7 -- 200 GeV becomes increasingly negative with decreasing collision energy. The overall significance of observing negative net-proton $C_6/C_2$ (0-40\% centrality) in more than half of the collision energies in the range $\sqrt{s_{NN}}$ = 7.7 –- 200 GeV is found to be 1.7$\sigma$. This significance is obtained by randomly varying the data points at each energy within their respective total Gaussian uncertainties (statistical and systematic uncertainties added in quadrature), a million times (we call them trials). Then, the number of trials out a million, where at least five or more collision energies have negative $C_6/C_2$, was calculated. This probability is obtained to be 95.3522\%, which corresponds to a 1.7$\sigma$ effect.\\
\paragraph*{{\bf{Ordering of cumulant ratios:}}}

The measurements of proton cumulant ratios in Au+Au collisions at $\sqrt{s_{NN}}$ =  3 GeV show a reverse ordering compared to the lattice QCD calculation~\cite{Bazavov:2020bjn}, namely $C_3/C_1 < C_4/C_2 < C_5/C_1 < C_6/C_2$.
Using a statistical test, we found that the observed ordering at $\sqrt{s_{NN}}$ = 3 GeV does not follow the lattice QCD expectation with a 3.8$\sigma$ significance. This significance is obtained by randomly varying all the six cumulants $C_{n, n\leq6}$ at 3 GeV, simultaneously within the total Gaussian uncertainties (statistical and systematic uncertainties added in quadrature), a million times (we call them trials). Then, from each new set of cumulants, cumulant ratios were constructed and the lattice-QCD-predicted ordering was checked. The number of trials in which the lattice-QCD-predicted ordering was observed was found to be 65. Thus, the probability that the expected ordering was not followed is, $1-65/1000000.0 = 0.999935$, which corresponds to a 3.8$\sigma$ effect. \\
In the main text, we also report the probability of the measurements at higher collision energies, $\sqrt{s_{NN}}$ = 7 -- 200 GeV, showing a reverse ordering as seen in $\sqrt{s_{NN}}$ = 3 GeV. The statistical test described above was performed for each higher energy, and we counted the number of trails out of one million in which the reverse ordering was observed. This probability at various energies is tabulated in Table~\ref{tab1_signi}.
\begin{widetext}
	\begin{table*}[!htb]
	\caption{Probability (in \%) of observing a reverse ordering as shown by $\sqrt{s_{NN}}$ = 3 GeV data, at various higher collision energies}
\centering   
\begin{tabular}{|c|c|c|c|c|c|c|c|c|c|}
	\hline
	$\sqrt{s_{NN}}$ (GeV) & 7.7 & 11.5 & 14.5 & 19.6 & 27 & 39 & 54.4 & 62.4 & 200 \\
	\hline
	Probability (in \%)  & 0.858 & 2.5991 & 8.0209 & 0.1756 & 0.1424 & 0.6911 & 2.192 & 10.0739 & 2.0769 \\
	\hline
\end{tabular}
\label{tab1_signi}
	\end{table*}
\end{widetext}

\end{document}